# MATHEMATICAL MODELING
# OF BIOCHEMICAL PROCESS

## A MATHEMATICAL MODEL OF THE METABOLIC PROCESS OF ATHEROSCLEROSIS

*V. I. GRYTSAY*

*Bogolyubov Institute for Theoretical Physics, National Academy of Sciences of Ukraine, Kyiv;*
*e-mail: vgrytsay@bitp.kiev.ua*

*A mathematical model of the metabolic process of atherosclerosis is constructed. The functioning of the polyenzymatic prostacyclin-thromboxane system of blood and the influence of a level of "bad cholesterol", namely low density lipoproteins (LDL), on it are studied. With the help of the numerical experiment, we analyze the influence of the concentration of molecules of fat on hemostasis of blood in blood vessels. The kinetic curves for components of the system, phase-periodic bifurcation diagrams, attractors for various modes, and Poincaré cross-section and image of a strange attractor are constructed. The complete spectra of Lyapunov indices, divergencies, KS-entropies, predictability horizons, and Lyapunov dimensions of the fractality of strange attractors are calculated. Conclusions about the structural-functional connections, which determine the dependence of hemostasis of a circulatory system on the level of cholesterol in blood are drawn.*

K e y   w o r d s: *self-organization, hemostasis, chaos, circulatory system, prostacyclin-thromboxane system, Lyapunov indices, KS-entropy.*

In the present work with the help of a mathematical modeling, we continue the study of a prostacyclin-thromboxane system of blood. We will investigate how low density lipoproteins (LDL) influence the dynamics of this metabolic process. In the construction of equations (Eqs.) of our model and the determination of its parameters, we used the results obtained by Prof. S.D. Varfolomeev and Prof. A. T. Mevkh. Their book and the fruitful collaboration with them [1-3] allow Prof. V.P. Gachok and the author to obtain calculation results similar to the experimental ones in the case where the system is in a stable state of hemostasis [4-9], which state characterizes a healthy blood vessel. It is the ideal state, which is attained by synchronization of the systems of thrombosis and antithrombosis. The dynamical stationary equilibrium arises. The desynchronization of these systems results in the appearance of autooscillatory modes in the metabolic process of a prostacyclin-thromboxane system. If the stationary kinetics is broken so that the level of thromboxane increases, then the coagulability of blood grows as well, and the appearance of thrombosis becomes possible in the circulatory system. On the contrary, if the level of prostacyclin increases, then the coagulability of blood decreases, and hemophilia occurs. If the autooscillatory mode arises, then the appearance of a thrombus as a result of increased coagulability on some time interval and its abruption under a decrease of the coagulability in the following time interval are possible. The actions of external and internal factors induce various modes in the system.

The metabolic process of coagulability of blood is considered by the author of the article as an open nonlinear system. The study was conducted using methods of nonlinear dynamics.

The kinetic model [4-9] allowed us to trace the effect of various levels of activities of phospholipases and concentrations of prostacyclin and thromboxane on properties of a biosystem, to determine the role of the arachidonic acid exchange between thrombocytes and endothelium, to analyze the influence of parallel processes running with the participation of arachidonic acid on kinetics of changes and on stationary levels of prostanoids, and to find the structural-functional connections of self-organization in the biosystem.

We will modify the given model by adding four nonlinear Eqs.. The other parameters remain unmodified. Within the model, we will study the influence of concentration of "bad cholesterol" (LDL) on the metabolism of a hemostasis of blood vessels. The principal reason for its elevated level is high dietary fat content. The excessive content of fat in organism causes formation of atheromatous plaques. They are aggregates of LDL on internal walls of blood vessels, causing stenosis. The atheromatous plaques grow over time. As a result, blood circulation slows down, which creates a deficit of nutrients in tissues. In this case, the arteries become denser and gradually lose their elasticity; i.e., atherosclerosis develops [10-17].

Atherosclerosis does not appear instantly, but arises gradually during the whole life. The excess of LDL is accumulated on arterial walls and is chemically modified. The modified LDL then stimulate adhesion of the endothelial cells to monocytes and T-cells. In addition, the endothelial cells secrete chemokines, which entrap T-cells in a trap of intima. Macrophages and T-cells produce numerous mediators of inflammation such as cytokines and cell division signaling molecules. In addition, macrophages express waste receptors, which help them to absorb modified LDL. Macrophages absorb LDL, by filling themselves by drops of fat. These foamy macrophages loaded with fat and T-cells form fat strips, which are earlier manifestations of atherosclerotic plaques. Molecules participating in inflammation facilitate further growth of a plaque and formation of a fibrous capsule above the lipid core. Thus, the dynamics of the metabolic process of accumulation of cholesterol in the intima of an arterial wall has the autocatalytic character. The unusual accumulation of cholesterol occurs. Its amount depends on aggregated thrombocytes and oxidized lipoproids, which depend, in turn, on the concentration of cholesterol in blood. This can explain the sharp growth of cholesterol plaques and the unexpected appearance of atherosclerosis in a person with medium level of cholesterol in blood [10].

During person's life, the metabolic process of the circulatory system permanently adapts to the conditions of nutrition. Respectively, the amount of cholesterol in the intima of blood vessels varies. At a high level of cholesterol, atherosclerotic damage to walls of blood vessels occurs. Inflammation induces propagation of atherosclerosis under imbalanced nutrition, unhealthy life style, and associated pathological conditions. Inflammation of blood vessels and thrombosis result from the permanent shift of hemostasis to a critical state for an organism. Studies of the oscillatory dynamics of the given metabolic process will allow one to investigate the process of self-organization of the metabolic process of hemostasis of a circulatory system under changes in blood cholesterol level. The presented model can serve as an example of the self-organization in the open dissipative system of human organism. This will allow one to study the regularity of metabolic processes in human organism from a single physical viewpoint of synergetics.

**The mathematical model and methods of its study**

The general scheme of the hemostasis with regard for the entry of "bad cholesterol" into blood is presented in Fig. 1 [6, 10]. According to this scheme, we construct the mathematical model of the given metabolic process (1)-(12) [6, 10]:

$$\frac{dA_t}{dt} = \frac{k_5 S}{(1+S+R^2)(1+k_6 T_x)} -$$

$$+ k_p A_p - k_t A_t - \alpha_1 A_t,$$

$$- \frac{k_7 A_t E_1}{(1+A_t+k_1 T_x)(1+E_1)} +$$

$$+ k_p A_p - k_t A_t - \alpha_1 A_t, \quad (1)$$

$$\frac{dT_x}{dt} = \frac{k_7 A_t E_1}{(1+A_t+k_1 T_x)(1+E_1)} -$$

$$- \frac{k_8 T_x^4}{(k_9 + T_x^4)} - \alpha_2 T_x, \quad (2)$$

$$\frac{dA_p}{dt} = \frac{k_2 S R^2}{(1+S+k_3 A_p)(k_4+R^2)} - \frac{k_{10} A_p E_2}{(1+A_p)(1+E_2)} +$$

$$+ k_t A_t - k_p A_p - \alpha_3 A_p, \quad (3)$$

$$\frac{dP}{dt} = \frac{k_{10} A_p E_2}{(1+A_p)(1+E_2)} -$$

$$- \frac{k_{11} T_x^* P^4}{(1+T_x^*)(k_{12}+P^4)} - \alpha_4 P, \quad (4)$$

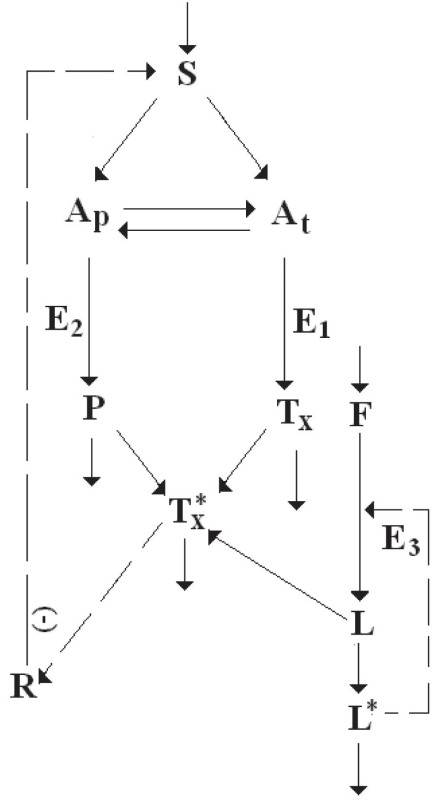

Fig. 1. General kinetic scheme of hemostasis

$$\frac{dE_1}{dt} = \frac{k_{13}A_T}{(1+A_T)(1+R^4)} -$$
$$-\frac{k_7 A_t E_1}{(1+A_t+k_1 T_x)(1+E_1)} - \alpha_5 E_1, \quad (5)$$

$$\frac{dE_2}{dt} = \frac{k_{15} A_p T_x^{*4}}{(k_{16}+A_p)(k_{17}+T_x^{*4})} -$$
$$-\frac{k_{10} A_p E_2}{(1+A_p)(1+E_2)} - \alpha_6 E_2, \quad (6)$$

$$\frac{dR}{dt} = k_{18}\frac{k_{19}+T_x^{*4}}{k_{20}+(T_x^*+k_{21}R)^4} - \alpha_7 R, \quad (7)$$

$$\frac{dT_x^*}{dt} = k_8\frac{L+T_x^4}{k_9+L+T_x^4} -$$
$$-\frac{k_{11} T_x^* P^4}{(1+T_x^*)(k_{12}+P^4)} - \alpha_8 T_x^*. \quad (8)$$

$$\frac{dF}{dt} = F_0 - l\frac{E_3}{1+E_3}\cdot\frac{F}{1+F+L}, \quad (9)$$

$$\frac{dL}{dt} = k\frac{E_3}{1+E_3}\cdot\frac{F}{1+F+L} - \mu\frac{LL^*}{1+L+L^*}, \quad (10)$$

$$\frac{dL^*}{dt} = \mu_1\frac{LL^*}{1+L+L^*} - \mu_0 L^*, \quad (11)$$

$$\frac{dE_3}{dt} = E_{30}L^*\frac{F}{1+F}\cdot\frac{N}{N+L} - \alpha_9 E_3. \quad (12)$$

The model includes the following collection of parameters [6]: $k = 4$; $k_1 = 3$; $k_2 = 1$; $k_3 = 5$; $k_4 = 10$; $k_5 = 2.1$; $k_6 = 5$; $k_7 = 2$; $k_8 = 1.5$; $k_9 = 5$; $k_{10} = 0.75$; $k_{11} = 0.3$; $k_{12} = 15$; $k_{13} = 0.75$; $k_{15} = 1$; $k_{16} = 0.5$; $k_{17} = 5$; $k_{18} = 5$; $k_{19} = 0.02$; $k_{20} = 25$; $k_{21} = 0.5$; $k_p = 0.1$; $k_t = 0.1$; $S = 2$; $\alpha_1 = 0.01$; $\alpha_2 = 0.01$; $\alpha_3 = 0.01$; $\alpha_4 = 0.173$; $\alpha_5 = 0.05$; $\alpha_6 = 0.07$; $\alpha_7 = 0.2$; $\alpha_8 = 0.0021$; $\alpha_9 = 0.2$; $F_0 = 0.01$; $l = 2$; $\mu = 4$; $\mu_0 = 0.437$; $\mu_1 = 2.3$; $E_{30} = 11$; $N = 0.05$.

The parameters of the system and time are dimensionless quantities [8].

These Eqs. (1)-(12) describe changes of concentrations of the dimensionless corresponding agents.

The input substances in a blood vessel are arachidonic acid $S$ and molecules of fat $F$, which are supplied into blood from the intestinal tract. The output agents of the system are aggregated thrombocytes $T_x^*$ and oxidized lipoproteins $L^*$, which are accumulated on internal walls of arteries. In the model we utilize the law of mass action and the kinetics of enzyme catalysis. The Eqs. involve the balance of masses of the intermediate products of reactions on separate stages of the metabolic process.

Eqs. (1) and (3) describe, respectively, changes in the concentrations of arachidonic acid $A_t$ and $A_p$ in thrombocytes and in endothelial cells of the vessel. These processes are affected by activity of the corresponding phospholipases. The coefficients $k_5$ and $k_2$ characterize, respectively, the rate of these processes. The accumulated arachidonic acid is transformed then by prostaglandin-$H$-synthase of thrombocytes $E_1$ and prostaglandin-$H$-synthase of prostacyclins $E_2$. Respectively, thromboxanes $T_x$ (2) and prostacyclins $P$ (4) are formed. The rate of these enzymatic processes is determined by the coefficients $k_7$ and $k_{10}$. The coefficients $k_t$ and $k_p$ characterize the exchange by arachidonic acid between thromboxanes and endothelial cells. Eqs. (5) and (6) describe, respectively,

77

changes of the concentrations of enzymes $E_1$ and $E_2$. The coefficients $k_{13}$ and $k_{15}$ determine the intensity of biosynthesis in thromboxanes and endothelial cells. The inactivation of these enzymes is guided by the corresponding terms with the coefficients $k_7$ and $k_{10}$ in Eqs. (2) and (4). Eqs. (1)-(6) satisfy completely the balance of masses in enzyme catalysis. Eq. (7) describes changes in the concentration of the controlling component, cyclic adenosine monophosphate (cAMP) $R$. Its presence in Eqs. (1) and (3) creates the negative feedback that affects the level of activity of phospholipases of thrombocytes $A_t$ and endothelial cells $A_p$. Equation (8) describes aggregation of thrombocytes $T_x^*$ and their dissipation under effect of prostacyclin. Its concentration depends also on blood LDL level. Eqs. (9)-(12) characterize the metabolic process of formation of LDL in blood and the accumulation of plaques on walls of the vessel. Fat molecules $F$ (9) are supplied by blood to arteries from liver and small intestine. Eqs. (9)-(10) describe the process of creation of "bad cholesterol" $L$ from fat. Its deposition on walls of blood vessels in the form of oxidized lipoproteids $L^*$ (plaques) is described by Eqs. (10)-(11). In the metabolic process, the positive feedback controlled by enzyme $E_3$ is formed (12). The accumulation of cholesterol in arteries and the growth of plaques cause thrombophilia. In this case, the lumen of an artery becomes narrower, i.e., stenosis develops. The above Eqs. involve also the dissipation of the corresponding substances at the expense of other metabolic processes and the flow of blood in an artery.

The given system is an open nonequilibrium one. Its study was carried out with the use of the theory of nonlinear differential Eqs. [18, 19] and the methods of mathematical modeling used earlier in [20-27].

### Results of Studies

The investigation of the mathematical model (1)-(12) has shown that, in addition to the ideal stationary modes of the metabolic process of the thrombosis-antithrombosis system [4-9], the model includes also autooscillatory modes. Depending on the rate of supply of molecules of fat $F_0$ to blood, the level of "bad cholesterol" $L$ varies. The metabolic process of hemostasis becomes unstable. The study of autooscillatory modes will enable us to comprehend the dynamics of the metabolic process and to reveal the structural-functional connections in this system.

In Fig. 2, we show a phase-parametric diagram of the system for $L(t)$ at a variation of $F_0$ in the corresponding intervals. In order to construct the phase-parametric diagrams, we have used the cutting method. In the phase space of a trajectory of the system, we place a cutting plane for the value of concentration of cyclic adenosine monophosphate (cAMP) $R = 1.77$. If the trajectory crosses this plane in some direction, we mark the value of chosen variable on the phase-parametric diagram ($L(t)$ in this case). Such choice is explained by the symmetry of oscillations of the given component relative to such point in multiply calculated earlier modes. For every value of $L$, we mark the intersection of the trajectory and this plane after the trajectory falls into the attractor. If a multifold periodic limiting cycle arises, we will see a number of points on the plane, which coincide in the period. If a deterministic chaos arises, the points of intersection are located chaotically.

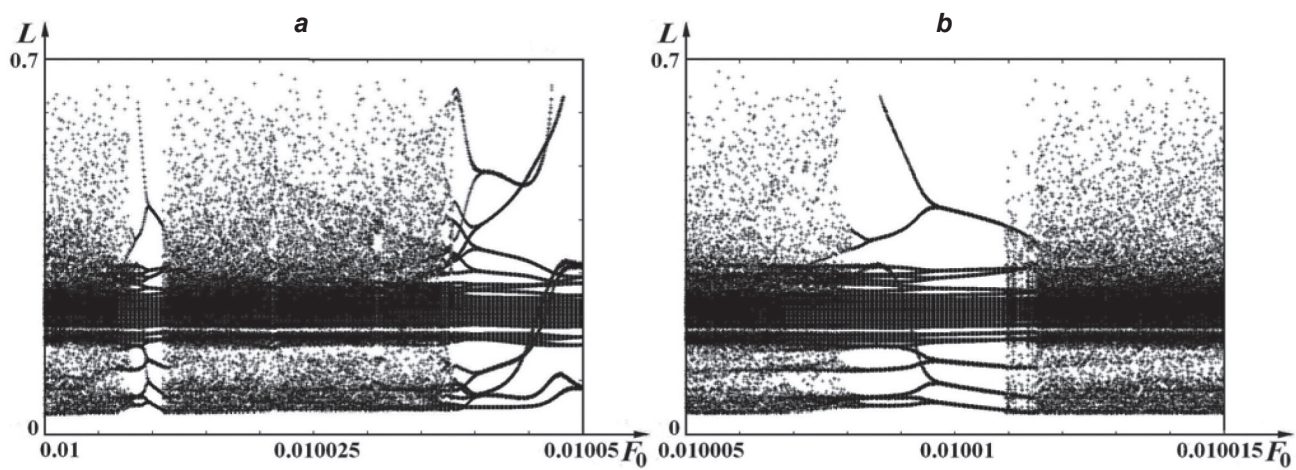

*Fig. 2. Phase-parametric diagram of the system for the variable L(t): a – $F_0 \in$ (0.01, 0.01005); b – $F_0 \in$ (0.010005, 0.010015)*

It is convenient to consider the phase-parametric diagram from right to left. As $F_0$ decreases (Fig. 2, *a*), we observe the successive transition to autoperiodic modes with higher multiplicity due to a cascade of bifurcations with the doubling of a period, until the chaotic mode is finally established due to the intermittence. Such scenario is shown in Fig. 3, *a-f*. The phase portraits of regular attractors transfer to a strange attractor. As the given parameter decreases further, the chaotic modes hold in the system. In the interval $F_0 \in (0.010007, 0.010012)$ (Fig. 2, *a*, *b*), chaos is destroyed, and the periodicity window arises. In this window, the transition from autoperiodic modes to chaotic ones occurs also as a result of the cascade of bifurcations with the doubling of a period by Feigenbaum's scenario. The transition finishes analogously to the previous scenario. Due to the intermittence, chaos is formed. The analogy of these scenarios indicates the fractal nature of the given cascades of bifurcations.

In Fig. 4, *a,b*, as an example, we show the projections of the strange attractor $2^x$ for $F_0 = 0.01$ in the planes $(L, F)$ and $(E_1, A_t)$. The obtained strange attractor is formed due to the funnel effect. An element of the phase volume of such attractor is stretched in some direction and contracts in other directions, by preserving its stability. Therefore, the mixing of trajectories happens in narrow contracted regions of the phase space of a funnel, and the deterministic chaos arises.

For the given strange attractor in Fig. 5 *a*, *b*, *c*, we constructed the projection of the intersection with the plane $R = 1.77$ and the Poincaré image. The intersection plane was chosen so that the phase trajectory $R(t)$ crosses it the maximal number of times, as the given component decreases, without any touching of the intersection plane by the phase curve.

The obtained intersection points and the Poincaré image do not possess a geometric self-similarity. The number of points permanently increases with the duration of a numerical integration of the system. This demonstrates the chaoticity of the attractor and the impossibility of some reduction of the given complicated kinetic scheme of metabolic processes to a one-dimensional discrete approximation of the system under study.

In Fig. 6, we compare the kinetics of some components of the system in the periodic (1) and chaotic (2) modes.

Changes of the concentrations of fat and "bad cholesterol" in the chaotic metabolic mode of atherosclerosis are shown in Fig. 7. Such nonuniform change of LDL affects the thrombosis-antithrombosis system, by destroying a steady hemostasis of an artery. The balance between the amounts of cholesterol deposited in a blood vessel and that taken out with blood is violated. At certain point the amount of deposited cholesterol becomes larger. This favors the formation of plaques in a blood vessel. Thus, the appearance of atherosclerosis depends on the self-organization of the metabolic process in the thrombosis-antithrombosis system. Under the self-organization, blood vessels adapt to the conditions of nutrition. If a desynchronization of these processes occurs, the risk of atherosclerosis development becomes more significant.

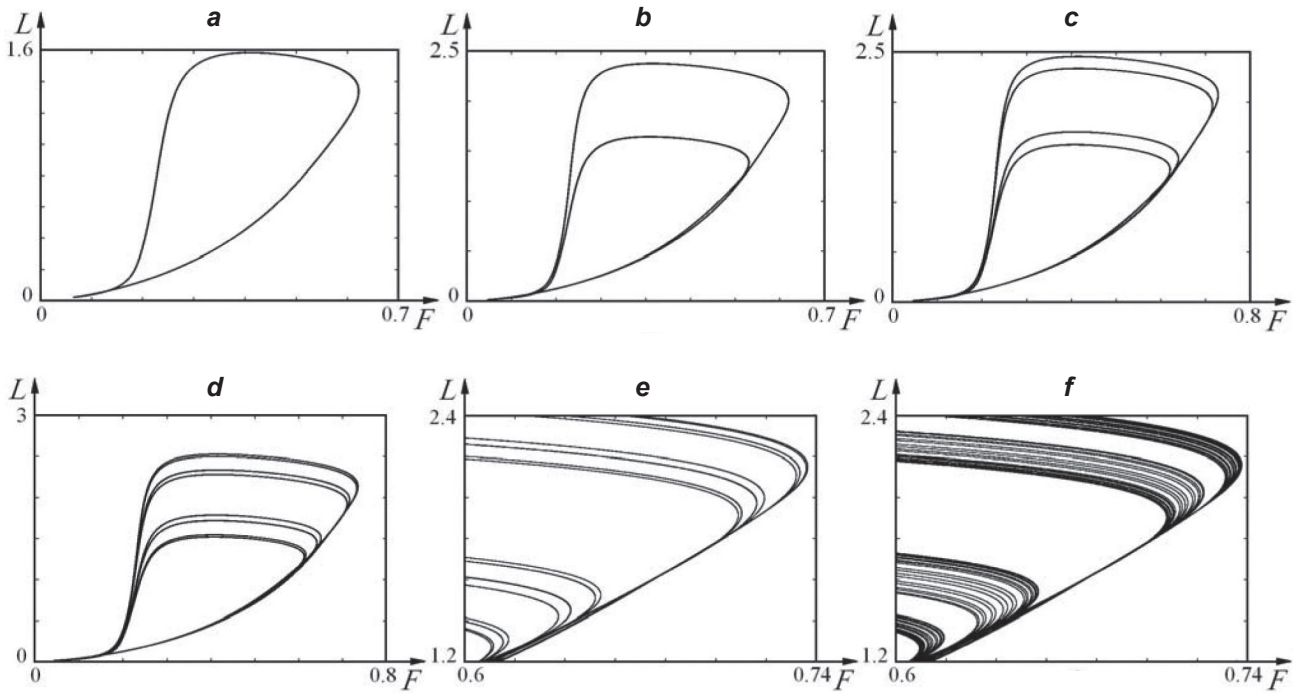

*Fig. 3. Projections of the phase portraits of attractors of the system in the plane (F, L): a – 1·$2^0$, for $F_0$ = 0.0102; b – 1·$2^1$, for $F_0$ = 0.01005; c – 1·$2^2$, for $F_0$ = 0.010045; d – 1·$2^4$, for $F_0$ = 0.0100383; e – 1·$2^8$, for $F_0$ = 0.010037; f – 1·$2^x$, for $F_0$ = 0.010036*

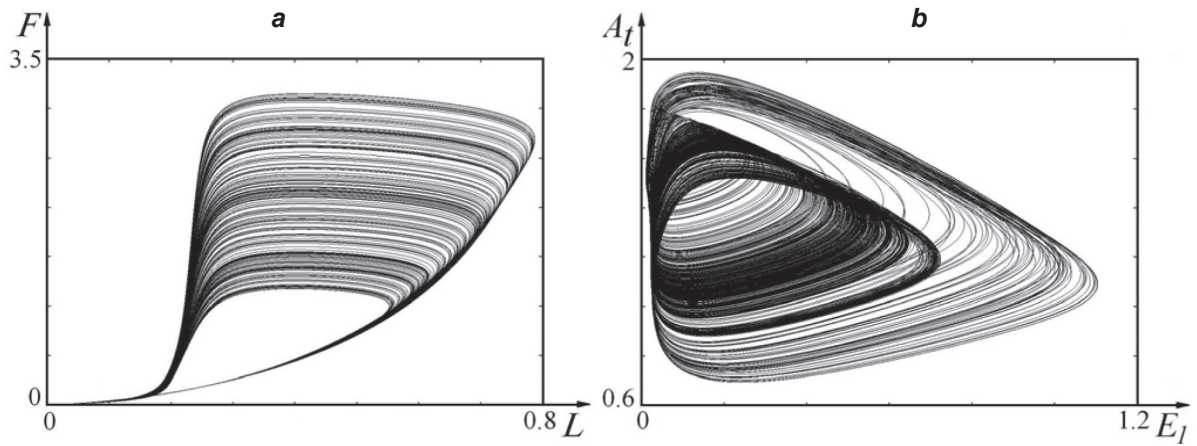

Fig. 4. Projections of phase portraits of strange attractor $2^x$, for $F_0 = 0.01$: a – in the plane (L, F), b – in the plane ($E_l$, $A_t$)

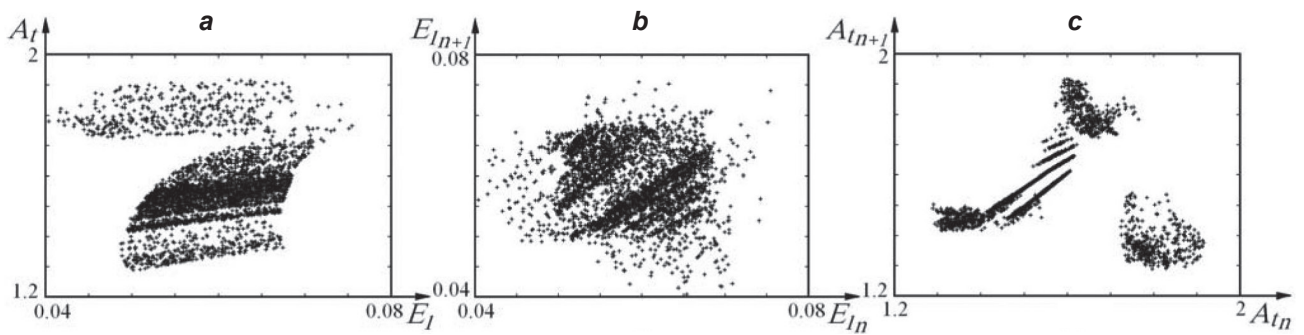

Fig. 5. Intersection projection (a) and the Poincaré images (b), (c) with the plane $R = 1.61$ for a strange attractor formed for $F_0 = 0.01$

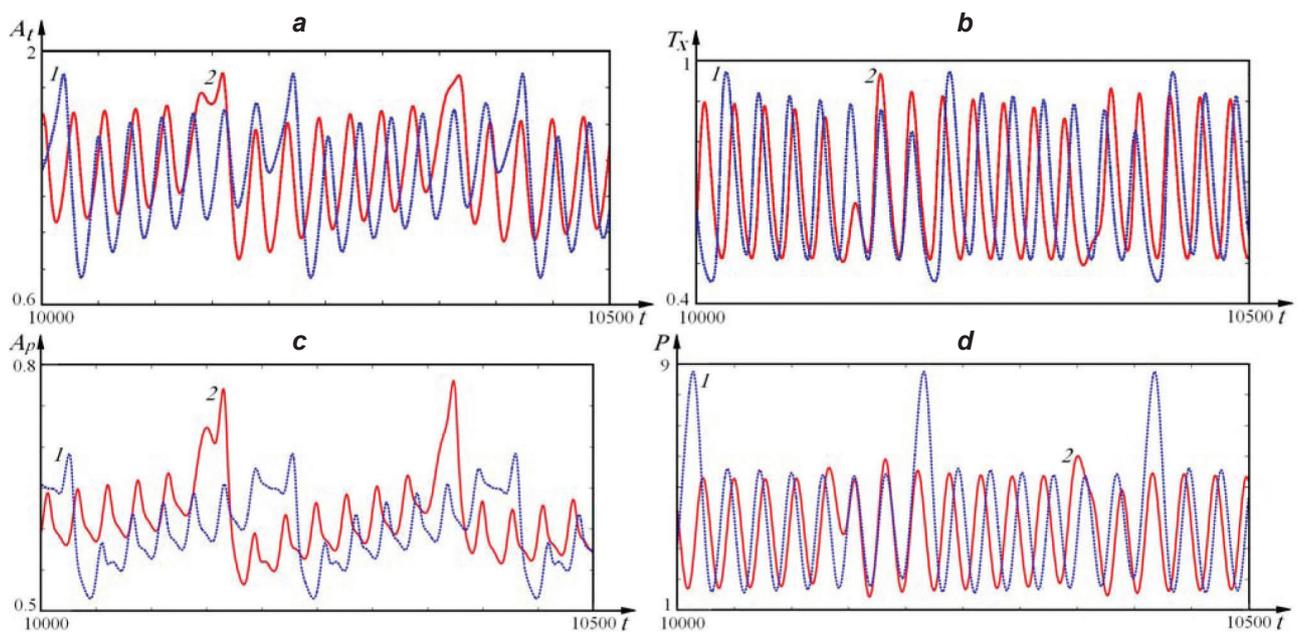

Fig. 6. Kinetic curves of the metabolic process of atherosclerosis during the running of the autoperiodic (1) ($F_0 = 0.0102$) and chaotic (2) ($F_0 = 0.01$) modes; a – $A_t$; b – $T_x$; c – $A_p$; d – P

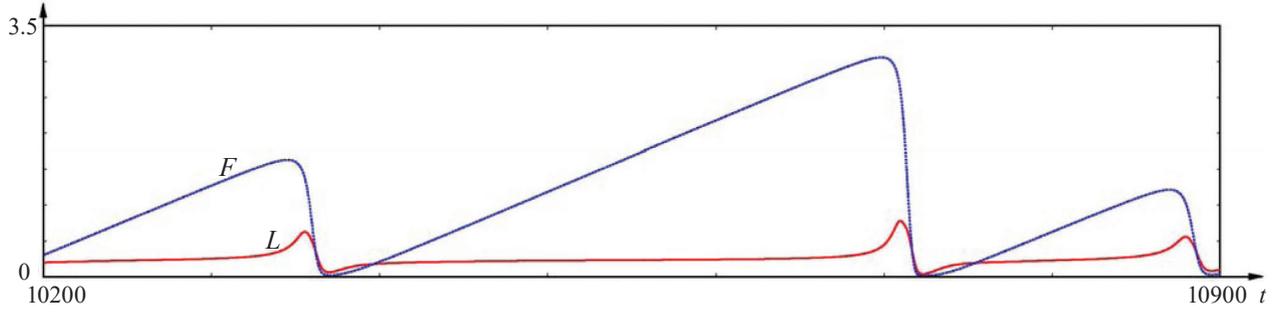

*Fig. 7. Kinetic curves of variations of the concentrations of F and L in the chaotic metabolic process of atherosclerosis for $F_0 = 0.01$*

For the unique identification of the type of the obtained attractors and for the determination of their stability for various values of parameter $F_0$, we calculated the complete spectra of Lyapunov indices and their sum: $\Lambda = \sum_{j=1}^{12} \lambda_j$. The calculation was carried out by Benettin's algorithm with the orthogonalization of the vectors of perturbations by the Gram-Schmidt method [18].

Using the Pesin theorem [28] and the values of Lyapunov indices, we calculated also the KS-entropy (Kolmogorov-Sinai entropy) $h$ and the Lyapunov index "predictability horizon" $t_{min}$ [29]. The Lyapunov dimension $D_{Fr}$ of the fractality of strange attractors was found by the Kaplan-Yorke formula [30, 31]:

$$D_{Fr} = m + \frac{\sum_{i=1}^{m} \lambda_i}{|\lambda_{m+1}|}.$$

Below, as an example for comparison, we present some results of calculations of the mentioned indices.

For $F_0 = 0.01$, the strange attractor $1 \cdot 2^x$ arises.

$\lambda_1 - \lambda_{16}$: 0.00192, 0.000197, -0.00432, -0.05045, -0.08331, -0.10046, -0.10209, -0.20994, -0.21184, -0.42267, -0.42267, -0.42267.

$\Lambda = -2.02831$; $h = 0.00192$; $t_{min} = 520.83$; $D_{Fr} = 2.44$.

For $F_0 = 0.010005$, the strange attractor $1 \cdot 2^x$ arises.

$\lambda_1 - \lambda_{16}$: 0.00172, 0.00018, -0.00370, -0.05334, -0.08149, -0.09988, -0.10109, -0.20985, -0.21308, -0.42381, -0.42381, -0.42381.

$\Lambda = -2.03197$; $h = 0.00172$; $t_{min} = 581.40$; $D_{Fr} = 2.46$.

For $F_0 = 0.01001$, the regular attractor $1 \cdot 2^0$ arises.

$\lambda_1 - \lambda_{16}$: 0.00004, -0.00044, -0.00155, -0.05447, -0.07744, -0.10082, -0.10476, -0.20904, -0.21267, -0.42695, -0.42695, -0.42695. $\Lambda = -2.04199$.

For $F_0 = 0.010015$, the strange attractor $1 \cdot 2^x$ arises.

$\lambda_1 - \lambda_{16}$: 0.00175, 0.00013, -0.00322, -0.05278, -0.08373, -0.09726, -0.10280, -0.21050, -0.21220, -0.42382, -0.42382, -0.42382.

$\Lambda = -2.03207$; $h = 0.00175$; $t_{min} = 571.43$; $D_{Fr} = 2.54$.

For $F_0 = 0.01002$, the strange attractor $1 \cdot 2^x$ arises.

$\lambda_1 - \lambda_{16}$: 0.00150, 0.00016, -0.00296, -0.05170, -0.08509, -0.09718, -0.10317, -0.21069, -0.21162, -0.42314, -0.42314, -0.42314.

$\Lambda = -2.03017$; $h = 0.00150$; $t_{min} = 666.67$; $D_{Fr} = 2.51$.

For $F_0 = 0.010025$, the regular attractor with quasiperiodic cycle $\approx n \cdot 2^0$ arises.

$\lambda_1 - \lambda_{16}$: 0.00033, 0.00009, -0.00173, -0.0526, -0.08456, -0.09697, -0.10372, -0.21086, -0.21112, -0.42345, -0.42345, -0.42345. $\Lambda = -2.03147$.

For $F_0 = 0.01003$, the regular attractor with quasiperiodic cycle $\approx n \cdot 2^0$ arises.

$\lambda_1 - \lambda_{16}$: 0.00097, 0.00018, -0.00388, -0.05107, -0.08488, -0.09845, -0.10156, -0.21077, -0.21243, -0.42252, -0.42252, -0.42252. $\Lambda = -2.02944$.

For $F_0 = 0.010035$, the regular attractor with quasiperiodic cycle $\approx n \cdot 2^0$ arises.

$\lambda_1 - \lambda_{16}$: 0.00065, 0.00013, -0.00408, -0.05188, -0.08540, -0.09748, -0.10064, -0.21061, -0.21283, -0.42306, -0.42306, -0.42306. $\Lambda = -2.03131$.

For $F_0 = 0.01005$, the regular attractor with quasiperiodic cycle $\approx n \cdot 2^0$ arises.

$\lambda_1 - \lambda_{16}$: 0.00002, -0.00005, -0.00158, -0.05371, -0.08570, -0.09922, -0.09722, -0.20979, -0.21411, -0.42708, -0.42708, -.42708. $\Lambda = -2.04259$.

These results show the variety of geometric structures of the obtained attractors and the predictability of the metabolic process depending on the concentration of molecules of fat in blood and on the level of "bad cholesterol".

Calculating successively various strange attractors, we can find some regularity in the hierarchy of their chaotic behavior. Respectively, the variation of the given indices changes a geometric view of attractors of the system.

Autooscilations in the metabolic process of hemostasis of a blood vessel arise due to the interaction between thrombosis and antithrombosis systems of blood, which is regulated by the level of cyclic adenosine monophosphate. The presence of "bad cholesterol" in blood causes desynchronization of these systems and the appearance of chaotic modes in the metbolism of a hemostasis. *LDL* affects the binding of thrombocytes and deposits on the walls of blood vessels. This leads to the autocatalysis of cholesterol in blood.

Thus, the hemostasis under a change of the amount of cholesterol in blood characterizes the adaptation of the metabolic process of a blood vessel to these changes, by preserving its functionality in this case.

We have constructed a mathematical model of the process of atherosclerosis of a blood vessel. The mathematical model describes the metabolic process of the thrombosis-antithrombosis system based on the prostacyclin-thromboxane system of blood. We have studied how molecules of *LDL* affect the imbalance of this system. The autooscillatory modes determined with this model indicate a complicated internal dynamics of formation of the self-organization in a blood vessel, i.e. that of the homeostasis.

We have studied the dependence of autooscillatory modes on the concentration of fat in blood. Moreover, we determined the chaotic modes of strange attractors. During such modes, the imbalance between the amount of "bad cholesterol" deposited in a blood vessel and its removal from the system happens. This provokes the formation of plaques in an artery. It is shown that affects the binding of thrombocytes and deposits on walls of blood vessels. This causes the autocatalysis of cholesterol in blood and the increase of its level. The mathematical study of the obtained modes is performed. The phase-parametric diagram, kinetic curves, projection of phase portraits, and Poincaré cross-sections and images are constructed. The Lyapunov indices, divergencies, "predictability horizons," and Lyapunov dimensions of the fractality of strange attractors are calculated. These indices characterize the stability and structure of calculated attractors.

The obtained results clarify the metabolic process of hemostasis and to find the structural-functional connections affecting the appearance of atherosclerosis of blood vessels.

The work is supported by the project N 0113U001093 of the National Academy of Sciences of Ukraine.

## МАТЕМАТИЧНА МОДЕЛЬ МЕТАБОЛІЧНОГО ПРОЦЕСУ АТЕРОСКЛЕРОЗУ

*В. Й. Грицай*

Інститут теоретичної фізики
ім. М. М. Боголюбова НАН України, Київ,
e-mail: vgrytsay@bitp.kiev.ua

Побудована математична модель метаболічного процесу атеросклерозу. Досліджується функціонування полієнзимної простациклін-тромбоксанової системи крові та вплив на неї рівня «поганого холестерина» – ліпопротеїнів низької щільності (LDL). За допомогою чисельного експерименту досліджується вплив рівня концентрації молекул жиру на гемостаз крові в кровоносних судинах. Побудовані кінетичні криві компонентів системи, фазоперіодичні біфуркаційні діаграми, атрактори різних режимів, переріз і відображення Пуанкаре дивного атрактора. Розраховані повні спектри показників Ляпунова, дивергенції, KC-ентропії, горизонти передбачуваності і ляпуновські розмірності фрактальності дивних атракторів. Зроблено висновки про структурно-функціональні зв'язки, що визначають залежність гемостазу кровеносної системи від рівня холестеролу в крові.

К л ю ч о в і  с л о в а: самоорганізація, гемостаз, хаос, кровеносна система, простациклін-тромбоксанова система, показники Ляпунова, КС-ентропія.

## МАТЕМАТИЧЕСКАЯ МОДЕЛЬ МЕТАБОЛИЧЕСКОГО ПРОЦЕССА АТЕРОСКЛЕРОЗА

*В. И. Грицай*

Институт теоретической физики
им. Н. Н Боголюбова НАН Украины, Киев,
e-mail: vgrytsay@bitp.kiev.ua

Построена математическая модель метаболического процесса атеросклероза. Исследуется функционирование полиэнзимной простациклин-тромбоксановой системы крови и влияние на нее уровня «плохого холестерина» – липопротеинов низкой плотности (LDL). При помощи численного эксперимента исследуется влияние уровня концентрации молекул жира на гемостаз крови в кровеносных сосудах. Построены кинетические кривые компонентов системы, фазопериодические бифуркационные диаграммы, аттракторы разных режимов, сечение и отображение Пуанкаре странного атрактора. Рассчитаны полные спектры показателей Ляпунова, дивергенции, КС-энтропии, горизонты прогнозируемости и ляпуновские размерности фрактальности странных атракторов. Сделаны выводы о структурно-функциональных связях, которые определяют зависимость гемостаза кровеносной системы от уровня холестерола в крови.

Ключевые слова: самоорганизация, гемостаз, хаос, кровеносная система, простациклин-тромбоксановая система, показатели Ляпунова, КС-энтропия.

### References

1. Varfolomeev S.D., Mevkh A.T. Prostaglandins – Molecular Bioregulators. M.: Moscow University, 1985. 308 p. (In Russian).
2. Varfolomeev SD, Mevkh AT, Gachok VP. Kinetic model of a multienzyme system of blood prostanoid synthesis. I. Mechanism of stabilization of thromboxane and prostacyclin levels. *Mol Biol.* (Mosk). 1986; 20(4): 957-967. (In Russian).
3. Varfolomeev SD, Gachok VP, Mevkh AT. Kinetic behavior of the multienzyme system of blood prostanoid synthesis. *Biosystems.* 1986; 19(1): 45-54.
4. Grytsay VI. The conditions of the self-organization in the multienzyme prostacyclin-thromboxane system. *Visn Kyiv Univ.* 2002; (3): 372-376.
5. Grytsay VI, Gachok VP. The modes of self-organization in prostacyclin-thromboxane system. *Visn Kyiv Univ.* 2002; (4): 365-370.
6. Grytsay VI, Gachok VP. Ordered structures in the mathematical system of prostacyclin and thromboxane model. *Visn Kyiv Univ. Ser. Fiz.-Mat. Nauk.* 2003; (1): 338-343.
7. Grytsay VI. Modeling of processes in the multienzyme prostacyclin and thromboxane system. *Visn Kyiv Univ.* 2003; (4): 379-384.
8. Gachok VP. Kinetics of Biochemical Processes. K.: Naukova Dumka, 1988. 224 p. (In Russian).
9. Gachok VP. Strange Attractors in Biosystems. K.: Naukova Dumka, 1989. 238 p. (In Russian).
10. Libby P. Atherosclerosis: the new view. *Sci Am.* 2002; 286(5): 46-55.
11. Verkhusha VV, Staroverov VM, Vrzhesh PV. The model of adhesive interaction of cells in the flow of a fluid. *Biol Membr.* 1994; 11(4): 437-450.
12. Baluda VP, Baluda MV, Deyanov II. Physiology of the System of Hemostasis. Moscow: Meditsina, 1995. 243 p.
13. Shitikova AS. Thrombocytic Hemostasis. St.-Petersburg, 2000. 225 p.
14. Libby P. Molecular bases of the acute coronary syndromes. *Circulation.* 1995; 91(11): 2844-2850.
15. Davies MJ. Stability and instability: two faces of coronary atherosclerosis. The Paul Dudley White Lecture 1995. *Circulation.* 1996; 94(8): 2013-2020.
16. Berliner J, Leitinger N, Watson A, Huber J, Fogelman A, Navab M. Oxidized lipids in atherogenesis: formation, destruction and action. *Thromb Haemost.* 1997; 78(1): 195-199.

17. Steinberg D. Low density lipoprotein oxidation and its pathobiological significance. *J Biol Chem.* 1997; 272(34): 20963-20966.
18. Anishchenko VS. Complex Oscillations in Simple Systems. Moscow, Nauka, 1990. 312 p. (In Russian).
19. Kuznetsov S.P. Dynamical Chaos. Moscow: Nauka, 2001. 296 p. (In Russian).
20. Kosterin SO, Miroshnychenko MS, Davydovska TL, Tsymbaliuk OV, Prylutskyy YuI. Phenomenologic mechanokinetic model of $Ca^{2+}$-dependent contraction-relaxation of smooth muscle. *Ukr Biokhim Zhurn.* 2001; 73(6): 138-142. (In Ukrainian).
21. Kosterin SO, Miroshnychenko MS, Prylutskyy YuI, Davydovska TL, Tsymbaliuk OV. Mathematical model of trans-sarcomere exchange of calcium ions and $Ca^{2+}$-dependent control of smooth muscle contractile activity. *Ukr Biokhim Zhurn.* 2002; 74(2): 128-133. (In Ukrainian).
22. Suprun AD, Prylutsky YuI, Shut AM, Miroshnichenko MS. Towards a Dynamical Model of Skeletal Muscle. *Ukr J Phys.* 2003; 48(7): 704-707.
23. Prylutskyy YuI, Shut AM, Miroshnychenko MS, Suprun AD. Thermodynamic and Mechanical Properties of Skeletal Muscle Contraction. *Int J Thermophys.* 2005; 26(3): 827-835.
24. Kosterin SO, Prylutskyy YuI, Borysko PO, Miroshnychenko MS. Kinetic analysis of the influence of inverse effectors (inhibitors and activators) on enzymatic (transport) activity of proteins. *Ukr Biokhim Zhurn.* 2005; 77(1): 113-125. (In Ukrainian).
25. Suprun AD, Shut AM, Prylutskyy YuI. Simulation of the Hill Equation for Fiber Skeletal Muscle Contraction. *Ukr J Phys.* 2007; 52(10): 997-1000.
26. Grytsay VI, Musatenko IV. Self-organization and fractality in a metabolic processes of the Krebs cycle. *Ukr Biokhim Zhurn.* 2013; 85(5): 191-200.
27. Grytsay VI, Musatenko IV. Self-organization and chaos in the metabolism of a cell. *Biopolym Cell.* 2014; 30(5):404-409.
28. Pesin Ya.B. Characteristic Lyapunov indices and the ergodic theory. *Usp Mat Nauk.* 1977; 32(4): 55-112.
29. Kolmogorov AN. On the entropy per unit time as a metric invariant of automorphisms. *DAN SSSR.* 1959; 154: 754-755.
30. Kaplan JL, Yorke JA. The onset of chaos in a fluid flow model of Lorenz. *Ann N Y Acad Sci.* 1979; 316(1): 400-407.
31. Kaplan JL, Yorke JA. A chaotic behaviour of multidimensional differential equations, in: Functional Differential Equations of Fixed Points, edited by H. O. Peitgen, H. O. Walther, Springer, Berlin, 1979. P. 204-227.